\documentclass[aps,preprintnumbers,superscriptaddress,nofootinbib,twocolumn]{revtex4}

\usepackage{ucs}
\usepackage{graphicx}
\usepackage{amssymb} 
\usepackage{amsmath} 
\usepackage{amsthm}
\usepackage[usenames,dvipsnames]{color}
\usepackage{bbm} % 1|

	\newcommand{\ket}[1]{\vert  #1 \rangle}
	\newcommand{\bra}[1]{\langle #1 |}
	
	\newcommand{\proj}[2]{\ket{#1}\bra{#2}}
	\newcommand{\pure}[1]{\proj{#1}{#1}}
 
	\newcommand{\set}[1]{\left\{ #1 \right\}}
	\newcommand{\avg}[1]{\langle #1 \rangle}

	\newcommand{\h}[1]{  {\cal H}_{#1}  }

	\newcommand{\one}{\! \! \uparrow}
	\newcommand{\zero}{\! \! \downarrow}
	
	\newcommand{\tr}{ \mbox{Tr}  }

	\newcommand{\id}{\mathbbm{1}}

	\newcommand{\hmin}{ H_{\min} }
	\newcommand{\hmax}{ H_{\max} }

\newcommand*{\eps}{\varepsilon}
\newcommand*{\cM}{\mathcal{M}}

\theoremstyle{plain}
\newtheorem{theorem}{Theorem}
\newtheorem{lemma}{Lemma}
\newtheorem{corollary}{Corollary}

\theoremstyle{definition}
\newtheorem{definition}{Definition}
	\newcommand{\iid}{i.i.d.\  }

	\newcommand \eref[1]{Eq.~\ref{#1}}
	\newcommand \fref[1]{Fig.~\ref{#1}}

\begin{document}

\title{The thermodynamic meaning of negative entropy}
%\date{31$^{\mbox{st}}$ August 2010}

\author{L\' idia \surname{del Rio}}
%\email[]{renner@phys.ethz.ch}
\affiliation{Institute for Theoretical Physics, ETH Zurich, Switzerland.}

\author{Johan \surname{{\AA}berg}}
%\email[]{renner@phys.ethz.ch}
\affiliation{Institute for Theoretical Physics, ETH Zurich, Switzerland.}

\author{Renato \surname{Renner}}
%\email[]{renner@phys.ethz.ch}
\affiliation{Institute for Theoretical Physics, ETH Zurich, Switzerland.}

\author{Oscar \surname{Dahlsten}}
%\email[]{renner@phys.ethz.ch}
\affiliation{Institute for Theoretical Physics, ETH Zurich, Switzerland.}
\affiliation{Clarendon Laboratory, University of Oxford, United Kingdom.}
\affiliation{Centre for Quantum Technologies, National University of Singapore, Singapore.}

\author{Vlatko \surname{Vedral}}
%\email[]{renner@phys.ethz.ch}
\affiliation{Clarendon Laboratory, University of Oxford, United Kingdom.}
\affiliation{Centre for Quantum Technologies, National University of Singapore, Singapore.}

\begin{abstract}
Landauer's erasure principle exposes an intrinsic relation between thermodynamics and information theory:
the erasure of information stored in a system, $S$, requires an amount of work proportional to the entropy of that system.
This entropy, $H(S|O)$, depends on the information that a given observer, $O$, has about $S$, and the work necessary to erase a system may therefore vary for different observers.
Here, we consider a general setting where the information held by the observer may be quantum-mechanical, and show that an amount of work proportional to $H(S|O)$ is still sufficient to erase $S$.  
Since the entropy $H(S|O)$ can now become negative, erasing a system can result in a net gain of work (and a corresponding cooling of the environment).  
\end{abstract}

\maketitle

\section{Preliminaries}

\label{sec:pre}

Statistical mechanics and information theory have a long standing and intricate relation. 
A famous example of this connection is Landauer's erasure principle \cite{landauer61}, used to exorcise Maxwell's demon \cite{bennett82}. 
According to this principle, in order to perform irreversible operations on a system, like the erasure of a bit of information, we need to perform work on the system, which is dissipated as heat to the environment. 
The necessary amount of work is determined by our uncertainty about the system~|~the more we know about the system, the less it costs to `erase' it. 
This result suggests that the seemingly elusive concept of `information' is directly linked to a very concrete quantity, `work'. 
Here, we analyse the relation between thermodynamics and information in a world that is fundamentally quantum mechanical.

Quantum information theory has peculiar properties that cannot be found in its classical counterpart. 
One example is that one's uncertainty about a system, as measured by an entropy, can become negative~\cite{horodecki05}. 
This motivates the following question: when our uncertainty about a system is negative, can we \emph{gain} work by erasing the information stored in that system?
Our results show that this is indeed possible; inherently non-classical aspects of quantum information theory, like negative uncertainty, are at a fundamental level part of thermodynamics.

\subsection{Physics from an information-theoretic viewpoint}
\label{sec:infview}

Our knowledge about the state of physical systems is usually limited,  because the number of parameters that we can measure and store, as well as our precision, are finite.
A typical example is a gas: we cannot keep track of the state of each particle, but only of a few macroscopic parameters, such as the volume or pressure of the gas. 
Despite this restricted information, it is possible to make accurate predictions about the behavior of systems using tools of statistical mechanics~\cite{Jaynes57, Jaynesb57,  Penrose79}.

Information constraints can also result in different observers having considerably different knowledge about the same physical reality.
To illustrate this \emph{subjectivity of information}, consider an $n$-qubit system, $S$ (e.g., $n$ spin-$1/2$ particles). 
An observer, Alice, prepares the system in a known pure state. 
A second observer, Bob, does not know which state that is, but applies an energy measurement to the system.
If $S$ is degenerate, Bob remains ignorant about the exact state of the system.

A natural way to quantify the knowledge of these observers is to use entropy measures.
The \emph{entropy} of a system, $S$, given all the information available to a given observer, $O$, denoted by $H(S|O)$, increases with the uncertainty of the observer about the exact state of the system.\footnote{For
	concreteness, one may think of the von Neumann entropy, which
	for a system, $S$, in state $\rho$, is defined by $H(S)_\rho := - \tr(\rho \log_2 \rho)$.  
However, most of this section is valid for any reasonable entropy measure,
and our technical statements will use smooth min- and max-entropies~\cite{renner05}. 
These are generalizations of the von Neumann entropy, and reduce to the latter for certain `nicely behaved' distributions, e.g., in the thermodynamic limit (see Appendix~\ref{appendix:entropies} for details). The subscript in  $H(S)_\rho$ can be dropped if the state is clear from the context.
} 
In the case where $S$ is fully degenerate, the entropy of the system from the point of view of Alice is zero, $ H(S|A)  = 0 $, as she has complete knowledge of the state of the system. 
On the other hand, Bob has maximal entropy, $ H(S|B)  = n $, because he does not know in which of the $2^n$ possible states the system is.\footnote{
	The entropy of $S$ conditioned on the classical memory $O$, $H(S|O)$,
	can be defined as the expectation, taken over all states of the memory, $m_O$, of the entropy of $\rho^m$, the state of $S$ conditioned on $m_O$: 
	$H(S|O) := \mathbb{E}_m\bigl[ H(S)_{\rho^m}\bigr]$.
} 

This observer-dependence of entropy seems to contradict the traditional thermodynamics view, where entropy appears as a property of the system rather than of the observer. 
However, the two views can be reconciled by introducing a \emph{standard observer} who has access to a well-defined set of macroscopic parameters, but whose uncertainty about the state of the system is otherwise maximal~\cite{Jaynes57}.
The idea is that the knowledge of this standard observer corresponds, to good approximation, to the knowledge we typically have about large systems in realistic situations: in general, we do not know microscopic details such as the spin direction of individual particles, but only parameters like the energy of a system (in the above example, it would make sense to take Bob as the standard observer).
One may nevertheless ask whether the difference between the entropies $H(S|A)$ and $H(S|B)$ has any physical significance. As we shall see, this is indeed the case.

\subsection{Quantum knowledge}
\label{section:quantknow}

The observers we described require an internal memory to store the information they have about the system $S$ (for Alice this memory needs to be large enough to include a full description of the state of $S$, while Bob only stores the value of the energy). It often is implicitly assumed that this memory is \emph{classical}. 
We go beyond this classical scenario and consider observers who may have access to information about $S$ that is itself represented as the state of a \emph{quantum} system --- a quantum memory. 

To illustrate the effects of a quantum memory, let us consider a third observer, Quasimodo.
Quasimodo prepares each of the $n$ particles of $S$ such that it is maximally entangled with a corresponding qubit of his quantum memory, $Q$. 
Note that this quantum memory is at least as useful as the classical data held by Alice. In fact, the latter may be recovered by applying a measurement on Quasimodo's memory.

In order to quantify the uncertainty that Quasimodo has about $S$, we need entropy measures that account for the quantum-mechanical nature of the information he holds. 
In the field of quantum information, such measures are known as \emph{conditional entropies} and generalize classical conditional entropies.
The conditional von Neumann entropy can be written as a difference,
$
H(S|Q) = H({S Q}) - H({Q}) .
$\footnote{If $Q$ was classical, this expression would be equivalent to $H(S|Q) := \mathbb{E}_m\bigl[ H(S)_{\rho^m}\bigr]$, as before.}
Here, $H({S Q})$ denotes the von Neumann entropy of the joint state of the system, $S$, and the quantum memory, $Q$. Since this joint state is pure, its entropy is zero. 
On the other hand, the reduced state of the memory, $\rho_{Q}$, is fully mixed, which corresponds to the maximal entropy $H(Q) = n$. 
We therefore find that, for Quasimodo, the conditional entropy is negative, 
$
  H(S|Q) = - n . 
$
Such negative entropies cannot occur for purely classical observers like Alice and Bob. 

This raises the question of whether these `negative uncertainties' have any operational meaning. The answer is yes.
They can be used to quantify, for instance, the amount of entanglement needed to send a state to a receiver with side information, a task commonly referred to as
`state merging' \cite{horodecki05}.
Another example where negative conditional entropies play a crucial role was given recently in the context of Heisenberg's uncertainty principle.
The principle bounds the minimum uncertainty one has about the outcome of a measurement on a system, $S$, chosen from two complementary observables, e.g., a spin measured in the $X$ or $Z$ basis.\footnote{
	More precisely, in its formulation proposed by Deutsch~\cite{deutsch83} and Maassen and Uffink~\cite{maassen88}, the principle asserts that 
	$H(X|O) + H(Z|O) \geq \log_2 \frac{1}{c}$
, where $O$ is any \emph{classical} description of the initial state of $S$, and where $\log_2 \frac{1}{c} \geq 0$ is a measure for the non-commutativity of the observables $X$ and $Z$.}  
This bound is, however, violated if quantum information about the initial state of $S$ is available.  It was shown that this violation can be quantified by the negativity of the entropy of $S$ conditioned on the memory~\cite{berta10}.\footnote{
	In the generalized form where $O$ may be non-classical, the relation reads $H(X|O) + H(Z|O) \geq \log_2 \frac{1}{c} + H(S|O)$.}

In this work, we go one step further and establish a relation between a \emph{physical} quantity (namely the work necessary to `erase' the state of a  system) and the conditional entropy. Remarkably, the validity of this relation extends to the quantum regime and, in particular, yields a direct thermodynamical interpretation of negative conditional entropies.

\subsection{Information-work relation}
\label{sec:InWorkRel}

\begin{figure*}[t]
\center
\includegraphics[width= 0.7 \textwidth]{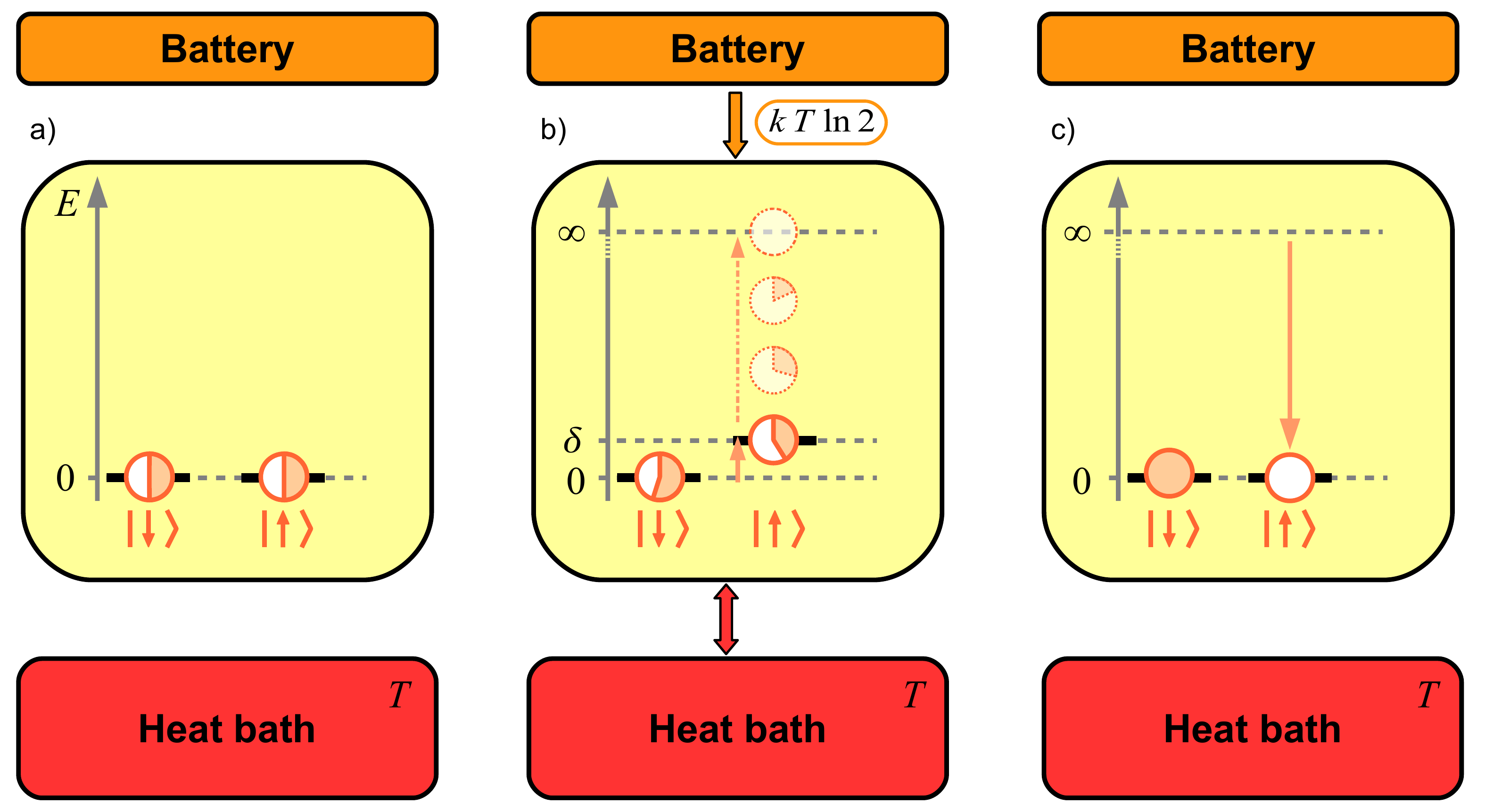}
\caption{
Erasing a fully mixed qubit.
$\left. a \right)$ 
	We start from a fully mixed state in a degenerate system. The filling of each circle represents the probability, $\avg{n_{\downarrow / \uparrow }}$, that the system is in the respective state.
$\left. b \right)$ 
	We couple the system to a heat bath at temperature $T$ and slowly raise the energy of state $\ket{\one}$. 
%Because the state is partially occupied, raising the level by a small amount $\delta$ costs in average $\avg{n} \delta $.
	Thermalized by the bath, the system equilibrates in a Gibbs state of temperature $T$. As the energy of $\ket{\one}$ increases, it becomes less occupied, according to $\avg{n_\uparrow}(E) = [1 + e^{E/ k T}]^{-1}$.
	We continue raising that level until it is empty. 
	The total cost of this operation is $\int_0^\infty \avg{n_\uparrow}(E) \ d E = k T \ln 2$.
$\left. c \right)$
	Finally, we isolate the system and lower the energy of state $\ket{\one}$. Since the state is empty, this operation is energy neutral.
}
\label{fig:erase1qubit}
\end{figure*}

In this section we illustrate Landauer's erasure principle and express it in terms of conditional entropies.
The process of \emph{erasing a system} is defined as taking the system to a pre-defined pure state, $\ket{0}$. 
Note that while erasing a system leads to the loss of information that could be encoded there,
%(for instance, if we knew that the system was previously in state $\ket{5}$)
it may also reduce our uncertainty about the system (if we did not know the previous state of the system, now we are sure that it is $\ket{0}$).

For a concrete example of how to erase a bit, consider a spin-$1/2$ particle 
exposed to a tunable magnetic field that can be adjusted to manipulate the energy of states $\ket{\zero}$ and $\ket{\one}$, according to a Hamiltonian like $\h{B} = j \vec{B} \cdot \vec{s}$.
Initially, the magnetic field is turned off, so the system is degenerate.
We define `erasing' as taking the spin to the pure state $\ket{0} := \ket{\zero}$.
Let us see how two different observers could do this.

Our first observer, Alice, knows that the particle is in a pure state, for instance $\ket{\one}$. In order to take the particle to  $\ket{\zero}$, she may apply a unitary operation, in this case a {\sc not} gate. This operation is reversible and has no energy cost.

The second observer, Bob, has no information about the initial state of the system, describing it as a fully mixed state, $\frac{\id}{ 2}$. 
One strategy he can follow to erase the bit is to couple the particle to a heat bath and slowly increase the magnetic field, raising the energy of state $\ket{\one}$ until its occupation decays, as shown in \fref{fig:erase1qubit}. This erasure process has an energy cost of $k T \ln 2$, where $T$ is the temperature of the bath and $k$ the Boltzmann constant.

More generally, in a hybrid setting where the system, $S$, may be quantum mechanical but the information about it is classical, the work, $W(S)$, required to erase $S$ is given by~\cite{bennett03} 
\begin{align} \label{eq:entropyworkrelstand}
  W(S) = H(S) \ k T \ln 2 \ .
\end{align}
Crucially, \eref{eq:entropyworkrelstand} relates work to a quantity that is, according to our discussion above, dependent on an observer. This apparent contradiction is resolved by reconsidering the meaning of $W(S)$. Note that in order to erase a system, we need to design an experimental setup that can, and in general \emph{must}, depend on the knowledge we have about it. Hence, rather than describing $W(S)$ simply as the `amount of work one needs to perform to erase system $S$', one may interpret it as the `amount of work that an observer with memory $O$ needs to erase $S$', and denote it by $W(S|O)$.
For an observer with a classical memory, $O_C$,\footnote{
	In the literature on Landauer's erasure principle the system to be erased is sometimes referred to as a `memory'. 
	However, for the sake of clarity we reserve the term `memory' exclusively for the observer's memory resources.} 
we have in general 
\begin{align} 
	\label{eq:entropyworkrel}
  W(S|O_C) = H(S|O_C) \ k T \ln 2 \ . 
\end{align}
We emphasize that this formula does not contradict \eref{eq:entropyworkrelstand}. Instead, it makes it explicit that the relevant quantities may depend on the knowledge of the observer and, in particular, may differ for different observers (in our example, Alice had zero entropy and consequently erased the bit at zero cost, while Bob had $H(S|B) = 1$ and had to perform work $k T \ln 2$; see also~\cite{oscar11} for a discussion).

Our contribution is to generalize this relation to the fully quantum case. We will be able to analyse what observers with quantum memories can do to erase a system, and how much that costs them.

\section{The general relation between information and work}

\label{sec:GenInWorkRel}

\begin{figure*}[t!]
\center
\includegraphics[width= \columnwidth]{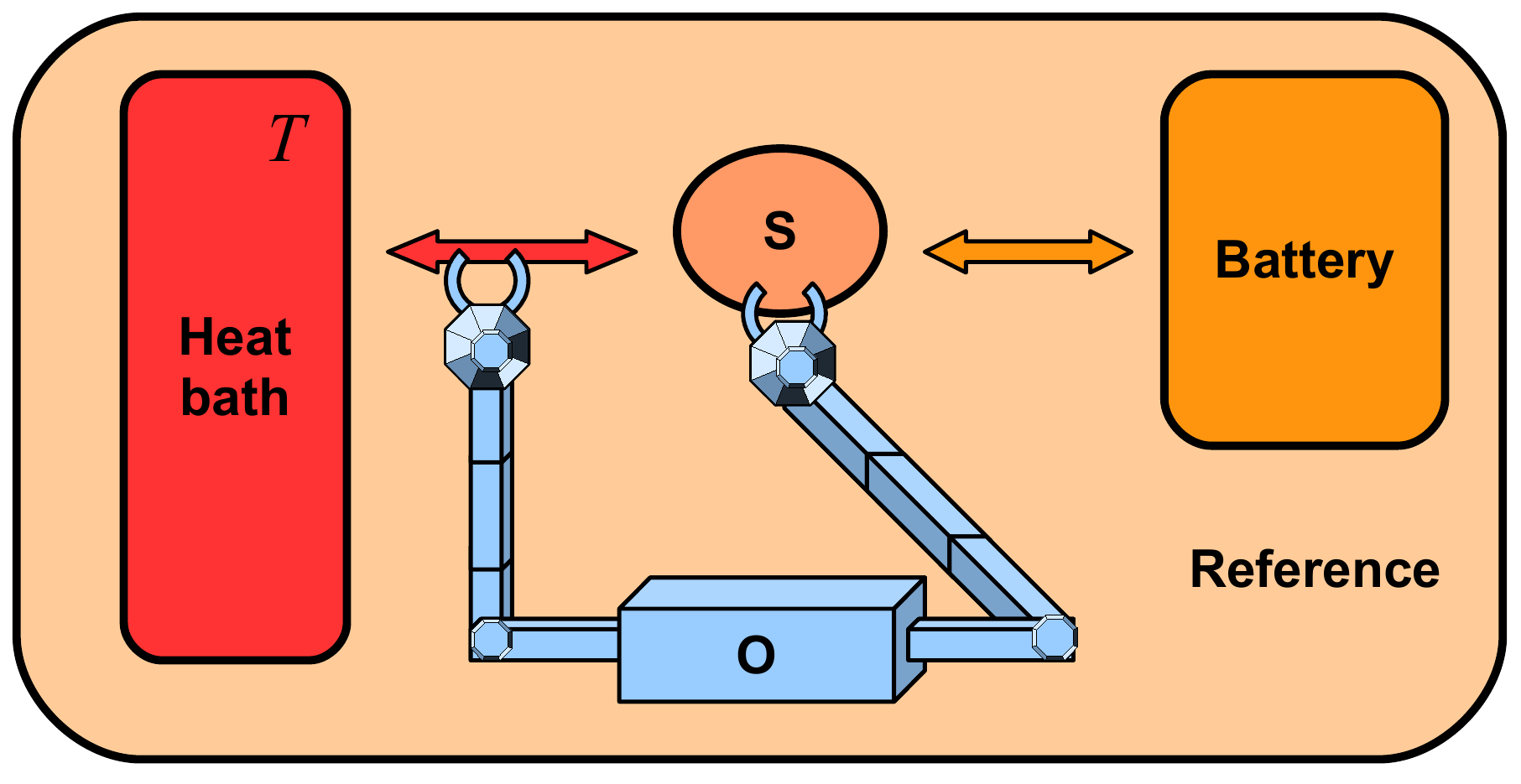}
\caption{Our setting: an observer, here represented by a machine with a quantum memory ($O$), will erase a system, $S$, using a heat bath at temperature $T$. 
The observer can store and withdraw energy from a battery. The rest of the universe is represented by the reference system.
}
\label{fig:setting}
\end{figure*}

In this section we state and explain our main result, a general relation between the work necessary to erase a system and the information one has about this system. 

Several approaches have been proposed in the past to formalize the idea of a thermal process and to study  erasure, work extraction and their relation to Maxwell's demon~\cite{landauer61, Szilard29, Shizume95, Piechocinska00, Janzing00, Parrondo01, Allahverdyan04, horodeckis05, Maroney09, Hilt10, oscar11}.
This has spurred a rather extensive literature (for overviews see~\cite{LeffRex90, LeffRex02, PlenioVitelli01, Maruyma09}) as well as debates (see, e.g.,~\cite{EarmanNorton98, EarmanNorton99,bub01,bennett03}).
Correlations and entanglement can affect erasure and work extraction, as has been noted by several authors. 
For instance, in~\cite{Oppenheim02} the system to be erased is bipartite and the observer is restricted to local operations and classical communication (LOCC); the difference between quantum and classical `demons' is addressed in~\cite{Zurek03}; see also~\cite{pebblesappreciationgroup10} for a discussion on `local' and `global' demons in the context of the thermodynamic arrow of time.

Here, we consider a setting as depicted in \fref{fig:setting}, where an observer, who has a quantum memory, $O$, tries to erase a system, $S$, using a heat bath at temperature $T$ and performing operations on $S$ and $O$ (which are not restricted to LOCC).
We assume that the initial Hamiltonian of  $S$ and $O$ is fully degenerate.
Details on the setting can be found in Appendix~\ref{appendix:setting}.

Since the memory $O$ is quantum mechanical, accessing it may in general change its content. Also, there is no reason why the memory would only contain information about
$S$; it could also carry information about other systems. 
Here we take a cautious position and require that those memory contents are kept intact in the erasure process.
Note that this requirement is crucial, since the contents may generally be  needed
for other purposes, e.g., if the erasure of $S$ is  part of a larger procedure.
As a simple example, suppose we erase system $S$, and later possibly would
like to erase another system $Z$. If the erasure of $S$ removed the
information about $Z$, the subsequent erasure of $Z$ could become
unnecessarily costly. 

In order to specify this memory preservation condition on a formal level, it is convenient to introduce a `reference system' $R$, which models all
systems other than $S$ that the memory can have information about.
To guarantee that the information about $R$ is unaltered, we assume that the
joint state of the memory and the reference, $\rho_{OR}$, is preserved by the erasure process and that system $R$ is not touched.

\subsection{A special case}
\label{sec:special}

\begin{figure*}[t]
\center
\includegraphics[width= 0.7 \textwidth]{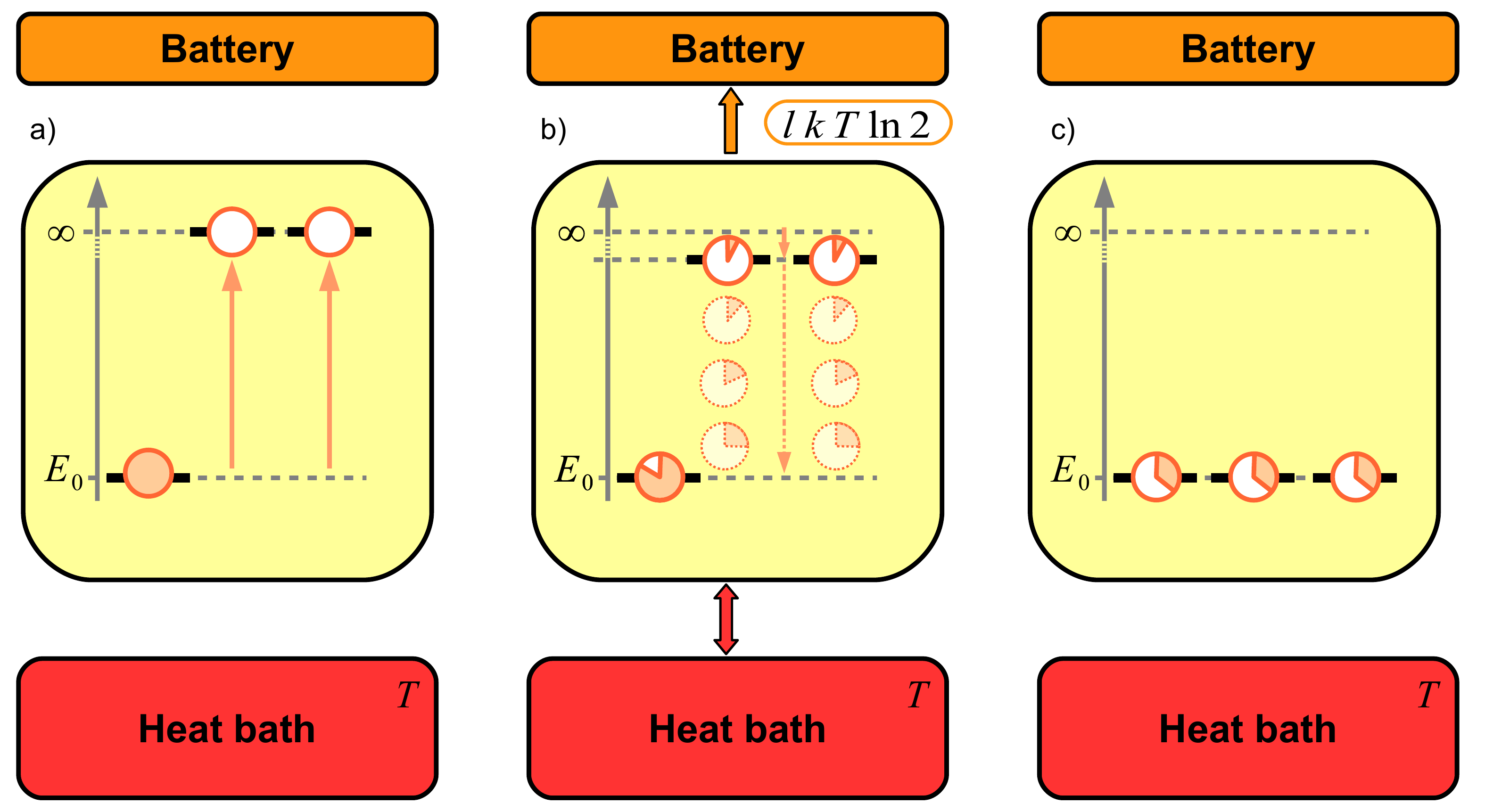}
\caption{Extracting work from a $\ell$-qubit system in a pure state. This process can be seen as the reverse of erasure (\fref{fig:erase1qubit}).
$\left. a \right)$ Only one state is occupied, at energy $E_0$; the energy of the empty levels is raised to a very high value at zero cost. 
$\left. b \right)$ We couple the system to the bath and slowly decrease the energy of the empty states. These will become gradually populated according to the Gibbs distribution. Lowering the partially occupied states results in energy gain of $\ell \ k T \ln 2$ in total. This energy is stored in the battery.  $\left. c \right)$ In the end of the procedure, the system is degenerate and fully mixed.
}
\label{fig:extracting}
\end{figure*}

The general idea of what an observer with a quantum memory can do to erase a system and \emph{gain} work in the process can be illustrated with a simple example.
Consider a single qubit system $S$, and an observer, Quasimodo, who has a memory formed by two qubits, $Q = Q_1 \otimes Q_2$. The first qubit is maximally entangled with $S$, in state $\ket{Q_1 S}$, while the second is maximally entangled with a qubit of the reference system, $R$, in state $\ket{Q_2 R}$.
Quasimodo will try to erase $S$ but keep his memory about $R$ intact, preserving the joint state 
$\rho_{QR} = \frac{\id_{Q_1}}{2}\otimes \pure{Q_2 R}$. Note that the reduced state of $Q_1$ is fully mixed, because $\ket{Q_1 S}$ is maximally entangled.

In a first step, Quasimodo uses the two-qubit pure state $\ket{Q_1 S}$ and a heat bath at temperature $T$ to extract work  $2 k T \ln 2$, as described in \fref{fig:extracting}. The system formed by $Q_1$ and $S$ is left in a fully mixed state. In particular, the reduced state of $Q_1$ is fully mixed, which implies that the joint state of the memory and the reference is still $\rho_{QR}$. 
Quasimodo then erases the fully mixed qubit $S$, like Bob did in Section~\ref{sec:InWorkRel}, performing work $k T \ln 2$. 
The net work gain of the whole procedure is $k T \ln 2$. 
Note that if Quasimodo had not preserved his memory and later wanted to erase $R$, he would have to perform unnecessary work.

This case illustrates how the relation between entropy and the work necessary to erase a system applies in a quantum scenario: Quasimodo had negative conditional entropy about $S$, $H(S|Q) = -1$, which resulted in negative work cost for erasure, $W(S|Q) = - k T \ln 2$. 

Naturally, the energy `gained' in this process comes from the heat bath.
As Quasimodo not only extracted work but also took $S$ to a pure
state, while leaving $\rho_{QR}$ intact, one may at first sight fear that
he has violated the second law of thermodynamics. 
This is, however, not the case, since those gains are balanced by
the reduction in correlations between $S$ and $Q$. 
In fact, the entropy of the global state, $H(QSR)$, increased, and
erasing $S$ made Quasimodo lose all the entanglement between his memory and $S$. 
His knowledge about the final state of $S$ is only classical~|~it can be expressed by a non-negative conditional entropy, $H(S|Q) = 0$. 
This prevents him from gaining more work if he erases $S$ again, using this process in a perpetual motion scheme.
The same observation also explains why a negative cost of erasure
would not enable Maxwell's demon to violate the second law.

\subsection{Single-shot erasure}
\label{sec:singleshot}

In general, the work required to erase a system
is a random variable, i.e., the cost of erasure 
may fluctuate each time it is performed.
Here we characterize a \emph{single instance} of erasure with a probabilistic statement, and in Section~\ref{sec:macro} we will consider the average work cost of erasure in a thermodynamic limit.

Theorem~\ref{thm:main} guarantees that the cost of erasing a system does not exceed a bound given in terms of the entropy of $S$ conditioned on $O$, except with a small probability. 

\begin{theorem}
\label{thm:main}
There exists a process to erase a system $S$, conditioned on a memory, $O$, and acting at temperature $T$, whose work cost satisfies
\begin{align}
\label{workmaxentr}
    W(S|O) \leq [\hmax^\eps(S|O) + \Delta]\ k\ T \ln 2, 
\end{align}
except with probability less than $\delta = \sqrt{2^{-\frac{\Delta}{2}} +12 \eps}$, $\forall \delta, \eps >0 $.
\end{theorem}

The quantity $\hmax^\eps(S|O)$ denotes the
$\eps$-smooth max-entropy of system $S$ conditioned on the quantum memory $O$,
a single-shot generalization of the von Neumann entropy~\cite{renner05}. In particular, as we shall see, this quantity reduces to the von Neumann entropy in a thermodynamic limit (we refer to Appendix~\ref{appendix:entropies} for definition and properties of smooth entropies).

The term $\Delta$ can be chosen to be small, and in the limit of large systems could be neglected.
For instance, to allow a maximum probability of failure of only $\delta = 3 \%$, one pays a price of approximately $20\ k T \ln 2$ in the work consumption of the process (in addition to the one dictated by the entropy).

Theorem~\ref{thm:main} implies that an observer with a quantum memory entangled with $S$ (i.e., with $\hmax^\eps(S|O) <0$) can erase the system with negative work cost, actually \emph{extracting} work in the process. 
Note that this is more general than the example of Section~\ref{sec:special}, where $S$ was, conveniently, maximally entangled with a part of the memory: Theorem~\ref{thm:main} implies that observers can make full use of the correlations between $S$ and $O$, even if those are not present in the neat form of maximally entangled qubits.

As a byproduct of the proof of Theorem~\ref{thm:main} we find an
analogous result for work extraction. 
The goal of this process is to extract work from a system, $S$, under the assumption that the memory is kept intact(while the final state of $S$ is arbitrary).
\begin{corollary}
\label{cor:workext}
Given an $n$-qubit system $S$ and a memory $O$, there exists a work extraction process acting at temperature $T$,  such that the extracted work satisfies   
\begin{align*}
    W_{e}(S|O) \geq [n - \hmax^\eps(S|O)  - \Delta]\ k\ T \ln 2 ,
\end{align*}
except with a  probability of at most $\delta=\sqrt{2^{-\frac{\Delta}{2}} +12 \eps}$, $\forall \delta, \eps >0 $.
\end{corollary}

\subsection{Thermodynamic limit}
\label{sec:macro}

\begin{figure*}[t]
\center
\includegraphics[width= 0.3 \textwidth]{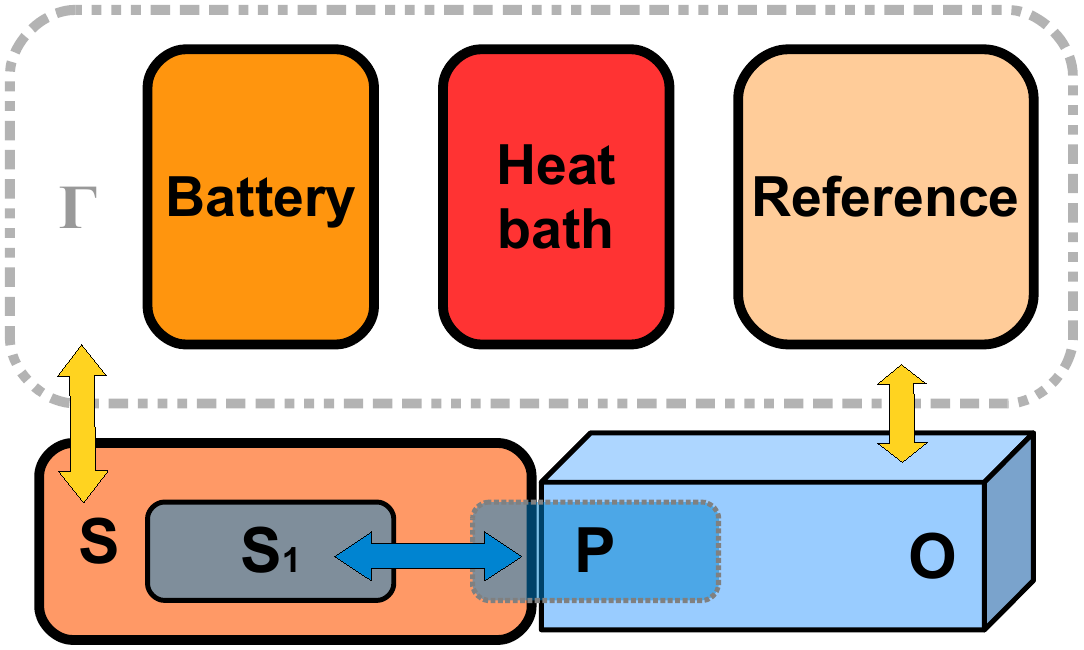}
\caption{Information compression, as used in the first step of our proof: a subsystem $S_1$ is decoupled from $\Gamma$. 
The size of $S_1$ decreases with the strength of the correlations between $S$ and $\Gamma$, and therefore increases with correlations between $S$ and the memory, $O$ (see Appendix~\ref{appendix:entropies}). 
Since the global state is pure, $S_1$ is purified by a system $P$ of equal size that belongs to the remaining systems, $S$ and $O$. The state of $S_1 \otimes P$ is fully entangled. The arrows symbolize correlations between the different systems.
}
\label{fig:decoupling}
\end{figure*}

We typically expect thermal fluctuations to disappear in macroscopic
systems. Theoretically, this is usually handled by
taking a thermodynamic limit, where we in some sense increase the size of
the system such that fluctuations are averaged away. 
In order to define a thermodynamic limit in our scenario, we imagine to perform the erasure on a large collection of independent systems. 

We define the \emph{work cost rate} of an erasure process as the average work cost of the process in this limit,
\begin{align*}
	\bar{w}(S|O) 
	= \lim_{n \to \infty}
		\frac{1}{n} 
		W(S^{\otimes n}|O^{\otimes n}).
\end{align*}
This quantity can be evaluated if we perform the erasure of many copies of a system. 
To understand the implications of our claim in such a situation, we use a well-known statement from information theory, the Asymptotic Equipartition Property (AEP)~\cite{ToCoRe09AEP}. The quantum version of this result essentially asserts that, for $n$-partite states that consist of many identical copies of the same single subsystem state, the smooth max-entropy converges towards the von Neumann entropy (see Appendix~\ref{appendix:entropies}).

The work cost rate can now be evaluated using Theorem~\ref{thm:main} combined with AEP, leading to the following result.

\begin{corollary}
	\label{thm:condVonNeumann}
	There exists a process to erase a system $S$, conditioned on a memory, $O$, and acting at temperature $T$, with work cost rate
	\begin{align*}
		\bar{w}(S|O)  \leq H(S|O) \ k T \ln 2.
	\end{align*}
\end{corollary}

\section{Outline of the proof}
\label{section:outline}

We prove our result by providing an explicit process that satisfies the bound of Theorem~\ref{thm:main}. We assume (without loss of generality) that $S$ is an $n$-qubit system.
The erasure process consists of three main steps:
\begin{enumerate}%[$\left. 1\right)$]
\item We manipulate $S$ in order to \emph{compress the correlations} between the memory and $S$ into a pure state of a subsystem of $S \otimes O$ that has approximately $n-\hmax(S|O)$ qubits. 
This state is maximally entangled between two subsystems of $S \otimes O$, like in the case of Quasimodo, from the example of Section~\ref{sec:special}.
\item We use that pure state to extract roughly $[n~-~\hmax(S|O)] \ k T \ln 2$ work ($k T \ln 2$ per qubit).
\item Finally, we erase system $S$, performing work $n \ k  T \ln 2$ (again, $k T \ln 2$ per qubit).
\end{enumerate}

We now describe these three steps in more detail, referring to technical proofs that can be found in the appendices when necessary.

In the first step, we show, using decoupling results~\cite{horodecki05,fred09} that, after an appropriate transformation, the first $\ell/2$ qubits of $S$ are almost (up to a probability determined by $\delta$) uncorrelated to the collection, $\Gamma$, of systems outside $S$ and $O$ (see Appendix~\ref{appendix:decoupling} for details), with 
\begin{align}
	\ell
		&\geq
		n - \hmax^\eps (S|O)
		+ 2 \log_2 (\delta^2 - 12 \eps).
\label{eq:boundsdecoupling}
\end{align}
These $\ell /2$ qubits form the subsystem $S_1$. As illustrated in \fref{fig:decoupling}, the fact that $S_1$ is decoupled from $\Gamma$ implies that there is an $(\ell/2)$-qubit subsystem, $P$, of $S \otimes O$ such that the state of $S_1 \otimes P$ is $\delta$-close to a pure, fully entangled state (details in Appendix~\ref{appendix:purification}).

In a second step, the observer extracts work $\ell\ k T \ln 2$ from the state of $S_1 \otimes P$ using a heat bath at temperature $T$, as described in \fref{fig:extracting} and Appendix \ref{appendix:extracting}. 
The system $S_1 \otimes P$ is left in a fully mixed state. 
Note that the state used was maximally entangled, so the reduced states of $S_1$ and $P$ were already fully mixed before this step. 
In particular, the part of the memory involved in work extraction
is not changed. The observer did not touch the memory before this second step and will not use it again, which implies that the reduced state of memory and reference,  $\rho_{OR}$, is preserved by the erasure process.
It is shown in  Appendix~\ref{appendix:extracting} that the probability of failure of work extraction is upper bounded by $\delta$. The work extraction process of Corollary~\ref{cor:workext} ends here.

In the last step of the erasure process, the observer uses energy from the battery to erase system $S$, as described in \fref{fig:erase1qubit}, performing work $n \ k T \ln 2$.
The work balance of whole process is $(\ell - n) \ k T \ln 2$. 
The logarithmic term in \eref{eq:boundsdecoupling} is usually negative, because we choose $\delta$ and $\eps$ to be small, so we can write the work consumption of the process as $W(S|O) \leq [\hmax^\eps (S|O)+ \Delta] k T \ln 2  $.

\section{Conclusions}
We have shown that conditional entropies, as measures of the uncertainty that
an observer has about a system, have a direct physical significance in
statistical mechanics. These results complement previous
findings that conditional entropies have an operational meaning within
information theory~\cite{horodecki05, berta10}. 
More specifically, we have introduced an erasure process that uses the quantum information that an observer has about a system to erase the latter. The work cost of this erasure process depends on conditional entropies, and 
a curious implication of our findings is that negative
entropies correspond to a negative work cost of erasure.
We have also seen that an observer with a quantum memory can extract twice as much work from a system as one with a classical memory.

The strengthened connection between information theory and statistical
mechanics may allow us to interchange concepts between the two areas.
An example is the proof of our results, as an essential part is played by \emph{decoupling},
which has shown to be a very powerful information theoretic primitive \cite{fred09, jonasquinn07}.
The following observation suggests that we  may also transfer ideas in the other direction. Intuitively, it appears rather clear that
observers cannot extract more work by locally processing data in their memory. Combined with our bounds for work extraction, this gives an alternative
`thermodynamic' derivation, as well as interpretation, of the data
processing inequality (also known as strong subadditivity) which, in
information theory, is a crucial and non-trivial result.

Our work can be related to \emph{discord}, a quantity originally
introduced in the context of open systems theory and decoherence \cite{Zurek00, Zurek02}, and also intensively studied in quantum information theory \cite{cavalcanti11,madhok11}. 
Discord quantifies the difference between the uncertainty about
a system, $S$, for an observer that possesses a quantum memory, $O_Q$, and
one that has
only a classical memory, $O_C$, obtained by performing a measurement on
$O_Q$, $\delta(S|O) = H(S|O_Q) - H(S|O_C)$. Similarly to
\cite{Zurek03,cornelio11}, our results suggest that $\delta(S|O)kT\ln 2$ can be
interpreted as the difference between the work cost of an erasure
procedure that makes full use of the quantum nature of the memory and a
process that is restricted to the classical properties of that memory.
In fact, since our relation between work and entropy is valid for a single instance of an erasure process, one may consider a generalized definition of discord based on the smooth max-entropy, which retains its operational meaning in the single-shot case.

\subsection{Applications}

Our result can also have implications on the fundamental limits of
computation.
Today, one of the major challenges to the miniaturization of circuitry
for
high-performance computing lies in the heat generation. With the
increased compactification, the heat generated per square unit of
circuitry is rapidly becoming difficult to handle. Although our
investigation certainly cannot help with the practical issues, it might
nevertheless be extended to a theory that provides the ultimate bounds on
dissipation. As is well known, computation per se can be made reversible
\cite{bennett73,fredkin82}. However, this comes
at the expense of keeping extra information about the computation in a
memory. Whenever we wish to erase a part of this memory, Landauer's
erasure principle dictates that this unavoidably comes at the cost of generating heat.

A very common scenario in a computation is that we wish to erase a part
of a memory, but
keep the rest of the memory intact. How much work do we need to dissipate
in order to do this? The naive answer would be that the cost is given by
the entropy solely of the part of the memory to be erased. 
However, our analysis shows that one can do better, namely that the required work is upper-bounded by a conditional entropy, which in general can be much smaller.

Note that our result requires almost perfect control of the quantum systems involved, and one may wonder why we should consider such a theoretical
idealization. As an analogue one can think of the Carnot cycle. Although
the ideal performance of the Carnot engine in many cases can be a
practically unattainable ideal limit, it nevertheless provides the
theoretical foundation in terms of which the performance of heat engines
can be gauged. 
Reversible computation together with the erasure principle
provides a similar ideal limit for minimally heat generating computation.

\acknowledgments{
We would like to thank Roger Colbeck, Sania Jevtic, Terry Rudolph, Tony Short and Will Matthews for discussions that helped shape the final form of this note. We address a special thanks to Charles H. Bennett for thorough conversations on the implications of this work.
We acknowledge support from the Swiss National Science Foundation (LdR, JA, RR, and OD, grant no.\ 200021-119868 and the NCCR QSIT), the Portuguese Funda\c{c}\~ao para a Ci\^{e}ncia e Tecnologia (LdR, grant no.\ SFRH/BD/43263/2008), the European Research Council (RR, grant no.\ 258932) and Singapore's National Research Foundation and Ministry of Education (VV). 
}

\subsection*{Regarding the appendix}

We emphasize that the Supplementary Information of the published version of this work~\cite{lidia11} has a more complete description of the formalism and proofs of our results than the present appendix. It can be accessed freely on the journal's website (unlike the Letter itself). We will update this appendix six months after publication, in December 2011.

%.......................... Appendix ....................

\appendix

\section{Formal setting}
\label{appendix:setting}

In this appendix we formalize the setting and the conditions for an erasure process that we use to derive Theorem~\ref{thm:main}.

\paragraph*{Setting:} our setting consists of a system $S$, a quantum memory, $O$, a heat bath at temperature $T$, a battery and a reference system, $R$ (\fref{fig:setting}), so that the initial global state is pure, and the Hamiltonian of the composed system $S\otimes O$ is fully degenerate.

\paragraph*{Allowed actions:}  the following physical processes on any subsystem, $X$, of $S \otimes O$ are allowed:
 unitary transformations on $X$;
 manipulation of the energy levels of $X$;
 coupling between $X$ and the heat bath or battery.
One may not perform any operations on the reference system.

\paragraph*{Erasure process:}
	in the setting described, a successful erasure process is one that erases system $S$ and preserves the joint state of the memory and the reference, $\rho_{OR}$. 
	A system is said to be \emph{erased} when it is in a pre-defined pure state.
	The \emph{work cost} of the process is defined as the difference between the initial and final charge of the battery.

Altering the energy of a state from $E_0$ to $E_0 + \Delta E$ has an average energy cost of $\avg{n} \Delta E$, where $\avg{n}$ is the probability that the system is in that state.
This energy can be withdrawn from a battery, modelled as follows.

\paragraph*{Battery.}
A battery is a system characterized by an energy value, $E$, called \emph{charge}, and the following operations:
	\begin{itemize}
		\item \emph{Withdrawing energy (performing work).} 
			If performing an operation on a system requires energy $\Delta E$, \emph{coupling} between the system and the battery is modelled by performing that operation and decreasing the charge of the battery by $\Delta E$.
		\item \emph{Storing energy (extracting work).}
			Conversely, if an operation on a system has a negative energy cost $\Delta E$, coupling the battery to the system and performing the operation results in an increment of $\Delta E$ of the charge of the battery.
\end{itemize}

\paragraph*{Heat bath.}
We assume that the heat bath is large enough to thermalize a system like $S$ without altering its own temperature. We model contact between a system and the heat bath by replacing the state of the system with a thermal Gibbs state of temperature $T$. 
Physically, this corresponds to letting the system be in contact with heat bath for long enough to thermalize. This condition does not imply that the state of the heat bath does not change~|~it does, losing or gaining the energy required to thermalize the system, but not enough to affect the temperature of the bath.

\section{Smooth entropies}
\label{appendix:entropies}

The main result, Theorem~\ref{thm:main}, relies on the smooth max-entropy, $\hmax^{\eps}$, as a measure to quantify uncertainty~\cite{renner05}. Smooth entropies have, so far, mainly been used in information theory, where they proved to be the relevant quantities to characterize information-processing tasks such as randomness or entanglement distillation, channel coding, data compression, or key distribution.

The formulation of the entropy-work relation in terms of the smooth max-entropy~|~rather than the more standard von Neumann entropy~|~has the advantage that the relation is valid independently of the structure of the underlying quantum states. A work-entropy relation involving the von Neumann entropy (Corollary~\ref{thm:condVonNeumann}) is obtained from this general result by introducing appropriate assumptions, as explained below. 

In the following, we briefly review the definition of smooth entropies and show how they are related to the von Neumann entropy. For a more detailed discussion of smooth entropies, their properties, and their information-theoretic significance, we refer to \cite{renes10, KoReSc09, datta09, renner05}.  

\subsection{Definition and properties}

Let $\rho = \rho_{S O}$ be the state of a bipartite system, consisting of subsystems $S$ and $O$.  The \emph{$\eps$-smooth max-entropy of $S$ conditioned on $O$} can be expressed in terms of the fidelity,\footnote{
	Note that the fidelity can be defined for arbitrary (not necessarily normalized) positive operators, $R$ and $S$, by 
	$F(R, S) := \| \sqrt{R} \sqrt{S} \|_1$, 
	where $\| \cdot \|_1$ is the $L_1$-norm.} 
$F$, as 
\begin{align*}
	\hmax^{\eps}(S|O)_{\rho} 
	:= \inf_{\rho'_{S O}} \sup_{\sigma_O} 
		\log_2 F(\rho'_{S O}, \id_{S} \otimes \sigma_O)^2 \ . 
\end{align*}
The supremum ranges over all density operators $\sigma_O$ on $O$. The infimum is taken over all (subnormalized) density operators $\rho'_{S O}$ that are $\eps$-close\footnote{
	Closeness is measured in terms of the \emph{purified distance}~\cite{ToCoRe09Duality}.}  
to $\rho_{S O}$, where $\eps \geq 0$ is the \emph{smoothness parameter}, which is usually chosen to be small but nonzero.

The proof of Theorem~\ref{thm:main} also involves the smooth min-entropy, which can be seen as the \emph{dual} of the smooth max-entropy, in the following sense. Consider a purification $\rho_{S O \Gamma}$ of the given bipartite state $\rho_{S O}$, with a purifying system $\Gamma$.  The \emph{$\eps$-smooth min-entropy of $S$ conditioned on $\Gamma$} then corresponds to the negative smooth max-entropy conditioned on $O$~\cite{KoReSc09,ToCoRe09Duality},
\begin{align}
  H_{\min}^\eps(S | \Gamma)_\rho = - H_{\max}^\eps(S | O)_{\rho} \ .
	\label{eq:duality}
\end{align}

Smooth entropies  have properties analogous to those of the von Neumann entropy. For example, for $\eps \to 0$, both $\hmin^\eps(S|O)_{\rho}$ and $\hmin^{\eps}(S|O)_{\rho}$ are $0$ if the reduced state on $S$ is pure, $1$ for a qubit $S$ that is fully mixed and uncorrelated to $O$, and $-1$ for a qubit $S$ that is maximally entangled with $O$. Furthermore, they satisfy a data-processing inequality. It asserts that the entropy of $S$ conditioned on $O$ can only increase if information is processed locally at $O$. Formally, 
\begin{align*}
  \hmax^\eps(S | O')_{\bar{\rho}} \geq \hmax^\eps(S | O)_{\rho},
\end{align*}
where $\bar{\rho} = \bar{\rho}_{S O'}$ is the state obtained from $\rho_{SO}$ when a completely positive map $\cM$ is applied on system $O$.

\subsection{Specialization to the von Neumann entropy}

For a bipartite quantum state $\rho_{S O}$, the \emph{von Neumann entropy of $S$ conditioned on $O$} is defined by 
$H(S | O)_{\rho} = H(\rho_{S O}) - H(\rho_O)$, 
where $H(\sigma)$ denotes the usual (non-conditional) von Neumann entropy of $\sigma$, i.e., $H(\sigma) = -\tr(\sigma \log_2 \sigma)$.  The conditional von Neumann entropy is always bounded by the smooth min- and max-entropies, 
\begin{align} 
  \lim_{\eps \to 0} H_{\min}^\eps(S | O)_{\rho} 
	&\leq H(S|O)_{\rho} \label{eq:vNbound} \\
	&\leq \lim_{\eps \to 0} H_{\max}^\eps(S|O)_{\rho} \ . \nonumber
\end{align}
In particular, if the smooth min- and max-entropies coincide, they are automatically equal to the von Neumann entropy.  Hence, under this condition, the smooth max-entropy occurring in Theorem~\ref{thm:main} can be replaced by the von Neumann entropy.  

A typical situation where \eref{eq:vNbound} holds (approximately) with equality is that of a large $n$-partite system with weakly correlated parts. In the limit when the correlations disappear, the state of the system is \emph{independent and identically distributed} (i.i.d.), i.e., of the form $\sigma^{\otimes n}$. Such states are common in information theory and physics~|~they arise, for instance, naturally for systems with sufficiently high symmetries (e.g., when a system is invariant under permutations of its $n$ parts~\cite{Renner07}).  One can show that the smooth min- and max-entropies converge for states of the form 
$\rho_{S^n O^n} = {\sigma_{S O}}^{\otimes n}$  \cite{ToCoRe09AEP}. Hence, by virtue of~\eref{eq:vNbound}, and using the fact that the von Neumann entropy is additive, one has, for any $\eps > 0$,
\begin{align} 
	\lim_{n \to \infty} &
		\frac{1}{n} H_{\max}^{\eps}(S^n | O^n)_{\sigma^{\otimes n}} \nonumber\\
	&= \lim_{n \to \infty} 
		\frac{1}{n} H_{\min}^{\eps}(S^n | O^n)_{\sigma^{\otimes n}} \nonumber\\
	&= H(S | O)_{\sigma} \ .
	\label{eq:AEP}
\end{align}
In other words, for \iid states, the work-entropy relation of Theorem~\ref{thm:main} asymptotically also holds for the von Neumann entropy. 

We note that \eref{eq:AEP} can be seen as a reformulation of the \emph{Asymptotic Equipartition Property}, which plays a crucial role in the area of information theory. There, operational quantities (such as the compression rate of a random source or the amount of randomness that can be distilled from a given source) are usually related to either the smooth min- or the smooth max-entropy. The widespread use of the von Neumann entropy in (text-book) information theory is therefore mainly a consequence of the fact that one typically considers \iid situations, such that \eref{eq:vNbound} holds with equality.

\section{Information Compression}
\label{appendix:compressing}

Here we address information compression, used in the first step of the erasure process; in particular, we prove the bound of \eref{eq:boundsdecoupling}, of Section $\ref{section:outline}$.

Information compression uses correlations between two systems, $S$ and $O$, as measured by an entropy measure, to create a pure state in a subsystem of $S\otimes O$, using only local reversible transformations on $S$.
In this result, we consider a global system $S \otimes O \otimes \Gamma$. In the context of our work, $S$ is the system the observer is trying to erase, $O$ the memory of the observer, and $\Gamma$ is formed by the battery, the heat bath and the reference system.

\begin{theorem}
\label{thm:compression}
Given a system $\Omega = S \otimes O \otimes \Gamma$ in a pure state, where $S$ is an $n$-qubit system, it is possible to create an $\ell$-qubit state of a subsystem of $S\otimes O$, with
\begin{align*}
	\ell
		&\geq
		n - \hmax^\varepsilon (S|O)
		+ 2 \log_2 (\delta^2 - 12 \varepsilon),
\end{align*}
that is $\delta$-close to a pure state, applying a local unitary transformation on $S$.
\end{theorem}

The last term is usually small. For instance, for $\delta= 0.003$ and $\varepsilon = 10^{-6}$, we have $2\log_2 (\delta^2- 12 \varepsilon) \approx - 20$. If the system $S$ is large (say $\approx 1000$ qubits), this logarithmic term can be neglected.

We will see later that the erasure process fails with maximum probability $\delta$. 
This means that allowing a probability of failure of $3\%$ has a cost of 10 qubits in the size of $S_1$, and results in an increase of $20 k T \ln 2$ in the work consumption of the erasure process (see Section~\ref{section:outline}).

The proof of Theorem~\ref{thm:compression} consists of two steps: first we will \emph{decouple} a subsystem $S_1 \subseteq S$, of $\ell/2$-qubits, from $\Gamma$. Then we will see that, since the global state is pure, $S_1$ is purified by a subsystem of $S \otimes O$ of the same dimension. The pure state created has a total of $\ell$ qubits.

\subsection{Decoupling}
\label{appendix:decoupling}

In this first step, we show that it is in general possible to identify a subsystem of $S$ that can be decoupled from $\Gamma$, according to the following definition.
\begin{definition}[Decoupling]
A system, $X$, is $\delta '$-decoupled from another system, $Y$, if their joint state is $\delta '$-close to a product state, 
\begin{align*}
	\delta \left( \rho_{XY}, \frac{\id_X}{|X|} \otimes \rho_Y \right) 
	\leq \delta
\end{align*}
where $\delta(\rho, \sigma) = \frac{1}{2} \| \rho - \sigma \|_1$ is the trace distance between two states.
\end{definition}

Lemma \ref{lemma:decoupling} will show that the size of the decoupled system depends on the correlations between $S$ and $O$, as measured by an entropy measure, the smooth conditional max-entropy, $\hmax^\eps(S|O)$. 
This result uses the procedure of decoupling, first introduced by \cite{horodecki05} and generalized by \cite{fred09}.

\begin{lemma}
\label{lemma:decoupling}
Given a system $\Omega = S \otimes O \otimes \Gamma$ in a pure state, where $S$ is an $n$-qubit system, it is possible to $\delta '$-decouple an $m$-qubit subsystem of $S$, $S_1$, from $\Gamma$. 
The maximum size of $S_1$ is given by 
\begin{align*}
	m
		&\geq \frac{n - \hmax^\varepsilon (S|O)}{2}
		+ \log_2 (2 \delta '- 12 \varepsilon).
\end{align*}
\end{lemma}

\begin{proof}

The decoupling results \cite{fred09, fred10} 
imply that the average distance between the state actually obtained after applying a unitary on $S$ and the desired, decoupled state, is given by
\begin{align}
	&\int_{U_S} 
		 \delta \left(   
		\tr_{S_2} \left(
		 [ U_S \otimes \id_{\Gamma} ]
		 \rho_{S\Gamma}^0  \right)
		, 
 		\frac{\id_{S_1}}{2^m}
		\otimes 
		\rho_{\Gamma}^0 
		\right)
		d U_S \nonumber \\
	& \quad \leq
		2^{-\frac{1}{2} \left( n - 2 m +2 \right)} 
			2^{- \frac{1}{2}\hmin^\varepsilon (S|\Gamma)_{\rho^0}}
		+6 \varepsilon .
\label{eq:firstbounddecoupling}
\end{align}
Here, the integral is taken over all unitary operations on system $S$, and $\hmin^\varepsilon (S|\Gamma)_{\rho^0}$ is the smooth conditional min-entropy of $S$, given the information that $\Gamma$ may provide about that system, before applying $U_S$. 
Since the bound of \eref{eq:firstbounddecoupling} applies to the average over all unitary operators, there is at least one fixed unitary, $U_S$, that respects it.
For an upper bound of $\delta '$ on the distance between the desired and the obtained states, we have
\begin{align}
	 &m  
		=
		\frac{n + \hmin^\varepsilon (S|\Gamma)}{2} 
		+ \log_2 (2 \delta ' - 12 \varepsilon).
	\label{eq:secondbounddecoupling}
\end{align}

The global state is pure, so one may use the duality relation between entropy measures, introduced in \eref{eq:duality} of Appendix \ref{appendix:entropies},
$
	\hmin^\varepsilon (S | \Gamma)_{\rho^0} 
		= -\hmax^\varepsilon (S | O)_{\rho^0}
$,
where the latter is the smooth conditional max-entropy of system $S$ given the memory. Inserting this to \eref{eq:secondbounddecoupling}, we obtain
\begin{align*}
	 &m  
		=
		\frac{n - \hmax^\varepsilon (S|O)}{2} 
		+ \log_2 (2 \delta ' - 12 \varepsilon).
\end{align*}

\end{proof}

It can be proved that the bound of Lemma \ref{lemma:decoupling} is optimal, i.e., that there is no unitary $U_S$ that allows us to decouple a system with more than $m$ qubits from $\Gamma$ \cite{fred10}.

\subsection{Purification}
\label{appendix:purification}
To complete the proof of Theorem \ref{thm:compression}, it remains to show that, given an $\frac{\ell}{2}$-qubit system $S_1$ decoupled from $\Gamma$, it is possible to find an $\ell$-qubit pure state in a subsystem of $S \otimes O$.
Note that the global state of $S \otimes O \otimes \Gamma$ is still in a pure state, for we have only applied a local unitary transformation on $S$.

\begin{lemma}
Consider a system $\Omega = ( S_1 \otimes S_2)  \otimes O \otimes \Gamma $ in a pure state, such that the $m$-qubit system $S_1$ is $\delta '$-decoupled from $\Gamma$, in a fully mixed state.

It is possible to find an $m$-qubit subsystem $P$ of $S_2 \otimes O$ that purifies the state of $S_1$ such that the joint state of $S_1 \otimes P$ is  $\sqrt{2 \delta}$-close to a fully entangled state.
\end{lemma}

\begin{proof}

In a first step we assume that the state of $S_1$ and $\Gamma$ if fully decoupled. We can expand it as
\begin{align*}
	\frac{\id_{S_1}}{2^m} \otimes \rho_\Gamma
	&= 2^{-m} \sum_k  \pure{k}_{S_1}
		\otimes \sum_i \lambda_i \ \pure{i}_\Gamma.
\end{align*} 
We can find systems $A_1$ and $A_2$ that purify $\rho_{S_1}$ and $\rho_{\Gamma}$. The composite system $A_1 \otimes A_2$ purifies $\rho_{S_1} \otimes \rho_{\Gamma}$,
\begin{align*}
	\ket{\phi }%_{S_1 A_1 \Gamma A_2}
		&= \ket{\phi '}_{S_1 A_1} \otimes \ket{\phi ''}_{\Gamma A_2}\\
		&= %\left( 
				2^{- \frac{m}{2}} \sum_{k} \ket{k}_{S_1} \ket{k}_{A_1} 
			%\right) 
			\otimes
			%\left(
				\sum_i \sqrt{\lambda_i}\ \ket{i}_{\Gamma} \ket{i}_{A_2} .
			%\right).
\end{align*}

The statement for $\delta '=0$ follows now from the fact that any two purifications of the same state are related by a unitary transformation on the purifying system. In particular, $P$ is given as the image of $A_1$ under this unitary. 
The claim for strictly positive $\delta '$ follows similarly, using Uhlmann's theorem and properties of the trace distance~\cite[Lem. 6]{ToCoRe09Duality}.  

\end{proof}

\section{Work extraction}
\label{appendix:extracting}

In this appendix we introduce in detail a process that allows us to extract energy from a heat bath and store it in a battery, using a pure state of a system $X$, as introduced in \fref{fig:extracting}.

\begin{theorem}
\label{thm:extracting}
Consider an $\ell$-qubit subsystem of $S\otimes O$, $X$, with a fully degenerate Hamiltonian, in a pure state. Using a heat bath at temperature $T$ and a battery, it is possible to extract exactly $\ell k T \ln 2$ work.  The system is left in a fully mixed state, and the final Hamiltonian is the same as the initial one.
\end{theorem}

\begin{proof}
Let $E_0$ be the energy of the initial state of $X$, $\ket{\phi_0}$, for a basis
$\set{\ket{\phi_i}}_i, i= 0, 1, \dots, N$.
We start by lifting the energy of all unoccupied states $\set{\ket{\phi_1} , \dots ,\ket{\phi_N}} $ to a high value, $E_1$. This can be done with no energy cost, because those states are empty (\fref{fig:extracting} $\left.a\right)$). 

Now we couple $X$ to the heat bath and let it thermalize; $X$ is taken to a Gibbs state of temperature $T$.
The probability that $X$ is in each of the states $\set{\ket{\phi_1} , \dots ,\ket{\phi_N}} $ is given by
$\avg{n} = \big[ N + e^{\beta (E_1- E_0)} \big]^{-1} $, where $\beta = (k T)^{-1}$. In total, the probability that the system is in one of the levels raised is 
$N \avg{n} =  \big[ 1 + e^{\beta (E_1- E_0) }/ N \big]^{-1} $.

We then couple $X$ to the battery and lower the energy of levels $\set{\ket{\phi_1} , \dots ,\ket{\phi_N}} $  by a small amount $\Delta$. Since those states were partially occupied, this operation gives us a small amount of energy, $ N \avg{n} \Delta$, that is stored in the battery (\fref{fig:extracting} $\left. b\right)$).

We wait for the system to thermalize again. Because levels $\set{\ket{\phi_1} , \dots ,\ket{\phi_N}} $  have slightly lower energy than before, they will become a little more populated, so the machine can extract a little more energy when it decreases the energy of the levels by another $\Delta$. 
The process is repeated until the energy of states $\set{\ket{\phi_1} , \dots ,\ket{\phi_N}} $  is lowered to $E_0$. At this point all $\set{\ket{\phi_i}}_i$ are degenerate again and the state of $X$ is fully mixed (\fref{fig:extracting} c).

In the quasistatic limit of $\Delta \rightarrow 0$ and $E_1 \rightarrow \infty$, this process allows us to extract a total amount of work of
\begin{align}
	&\lim_{E_1 \rightarrow \infty} 
	\int_{E_0}^{E_1}
		\frac{1}
			{1 + e^{\frac{\beta (E - E_0)}{N}}}
		dE\nonumber\\
	&\quad = \frac{\ln (N+1) }{\beta}
	= \log |X| \ k T \ln 2.
	\label{eq:extracting}
\end{align}

\end{proof}

The process described in Theorem~\ref{thm:extracting} takes a system from a pure to a fully mixed state, extracting some work in the process. By inverting the process (\fref{fig:erase1qubit}), one can bring a system from a fully mixed to a pure state~|~in other words, \emph{erase} the system.

\begin{corollary}
To erase an $\ell$-qubit system initially in a fully mixed state, using a heat bath at temperature $T$, it is sufficient to perform work $\ell k T \ln 2$.
\end{corollary}

When compressing information between the system and the memory, we allowed the state created to be at most $\delta$-distant from a pure state (Appendix \ref{appendix:compressing}). 
The following lemma shows how that affects the probability of failure of the work extraction procedure.

\begin{lemma}
If the process described in Theorem~\ref{thm:extracting} is applied to a state $\delta$-close to a pure state, it succeeds with probability at least $1-\delta$.
\end{lemma}

\begin{proof}
The probability that two states, $\rho$ and $\sigma$, of the same system can be distinguished in a one-shot approach using a physical process, such as a measurement after a reversible evolution, is given by ${\Pr}_{\max}(\rho, \sigma) = \frac{1}{2} [1 + \delta(\rho, \sigma)]$, where $ \delta(\rho, \sigma)$ is the trace distance between those states.

An example of a process to distinguish two states is the work extraction process described in Theorem~\ref{thm:extracting}.
If the process is applied to the expected pure state, $\sigma$, the probability of error is zero and the quantity of work extracted is $\ell\ k T \ln 2$. 
We denote the probability of failure of the work extraction process for an arbitrary state, $\rho$, by $p_\rho$.

If we are given one of the two states, $\sigma$ and $\rho$, at random, apply the work extraction process and obtain less than $\ell\ k T \ln 2$, we know that the state was $\rho$. This happens with probability $p_\rho/2$. 
In $(1 - p_\rho)/2$ of the cases, we are given $\rho$ and extract exactly $\ell\ k T \ln 2$, and with probability $1/2$ we had $\sigma$, extracting the same work, so our best guess if we obtain work $\ell\ k T \ln 2$ is to say we had state $\sigma$.
In total, we will be right with probability  $ \frac{1}{2} [1+p_\rho]$.

This guessing probability is upper bounded by ${\Pr}_{\max}(\rho, \sigma)$, so   $p_\rho \leq \delta(\sigma, \rho)$. 
Since we imposed a maximum distance $\delta$ between the pure state $\sigma$ and $\rho$, the probability of failure of the process is at most $\delta$.

\end{proof}

\section{Not-so-brief clarification}

The following notes concern the published version of
this manuscript~\cite{lidia11}, amplifying on some points and clarifying its relation
to earlier work on reversible
computing, Landauer's principle, and Maxwell's demon~\cite{bennett73,fredkin82,bennett82}.

\subsection*{Non-cyclic erasure}

Our paper deals with the isothermal work (at
temperature $T$) required by an observer $O$
to erase a system $S$, in other words restore it to a standard pure
state, $\ket{0}$. The observer may initially have knowledge about the system: $O$ may be classically correlated or even entangled with $S$.  We show that
the work cost rate required for erasure is given by
\begin{align} \label{eq:main} 
 \tilde{w}(S|O) = H(S|O) \ kT \ln 2,
\end{align}
where $H(S|O)$ is the conditional entropy of $S$ given $O$.

If the observer $O$ is classical, the work cost of erasure can be zero
(when $O$ has complete information on $S$) or positive (when $O$ has
partial or no information on $S$), but can never be negative.  More
generally, however, the observer $O$ may hold quantum information
(that cannot be represented by a classical value), and the study of
this more general situation is the main goal of our paper. In
particular, the initial correlations between $S$ and $O$ may be
quantum, and $H(S|O)$ can be negative. Eq.~\ref{eq:main} thus provides
an interpretation of this negative conditional entropy, namely that it
corresponds to a work yield, rather than a work cost, associated with
performing the erasure.

Landauer's principle was originally formulated as ``the cost of erasing an unknown bit is $k T \ln
2$'', and is generally taken to refer to a more limited situation, where
there are no correlations between $S$ and $O$, in other words where
the observer is entirely ignorant of the system being
erased. Conversely, a work yield of $kT \ln 2$ can be obtained
by quasi-statically allowing a qubit in an initial pure
state to randomize itself at temperature $T$ (Figure~3). In other words, one can gain work at the cost of losing all the initial information about the state of the qubit.  
Landauer's principle and its converse are generally seen as straightforward manifestations of the second law of thermodynamics, as applied to data-processing systems, and they can be applied in a cyclic fashion, e.g.\ to assess the work that Maxwell's demon needs to expend to clear its memory at the end of each cycle of operation.

In the case where the observer $O$ has non-trivial
  classical information about $S$, Landauer's principle may be refined
  to $
 \tilde{w}(S) = H(S)\ kT \ln 2,
$
where the entropy $H(S)$ is evaluated for
the state of $S$ conditioned on the knowledge held by $O$. Note that
this is consistent with Eq.~\ref{eq:main}, where the classical
knowledge of $O$ is made explicit, rather than taken implicitly
in the definition of the (conditional) state of $S$. 

However, in the general case of an observer who may hold quantum information about $S$, the implicit formulation of $\tilde{w}(S) = H(S)\ kT \ln 2$ is no longer possible, and conditional entropies are necessary to describe the knowledge of $O$ about $S$. In this sense, Eq.~\ref{eq:main} can be seen as a strict generalization of Landauer's principle to situations involving non-classical observers.
Let us reconsider the example of Quasimodo, who holds a quantum memory $Q$
maximally entangled with an $n$-qubit system $S$. It is one of the
central and celebrated features of quantum mechanics that, in this
entangled state, $S$ and $Q$ each appear maximally random, while the
joint $S Q$ system is in a pure state of zero entropy.  In other
words, we have $H(S Q) = 0$, $H(S) = H(Q) = n$, and therefore $H(S|Q)
= -n$. Eq.~\ref{eq:main} (with $Q$ taking the place of the
observer, $O$) therefore implies that the $n$-qubit system $S$ can be
erased with a negative work cost, i.e., a positive work yield, of $n\ 
kT \ln 2$.  
As before, erasure means that $S$ is brought to a
standard pure state $\ket{0}$, whereas $Q$ should remain unchanged, in the sense that the reduced density operator that describes the state of $Q$ should be the same before and after the erasure of $S$.
This follows from our information-preservation condition, which demands that erasure of $S$ must  not affect any other information held by the observer.

This erasure process might appear to risk violating the second law of
thermodynamics, for example by repeatedly allowing $S$ to randomize
itself, then extracting work as it is erased, in a cyclic fashion.
But in fact no such violation occurs, because the erasure process uses
up, and does not replace, the initial entanglement between $S$ and
$Q$, thereby preventing the cycle from repeating. 

During the non-cyclic erasure, the joint system $SQ$ evolves from a pure initial state to a final state with n bits of entropy, an entropy increase that can be harnessed to do  $n kT \ln 2$  of work, violating neither the second law nor the original unconditional form of Landauer's principle (which applies to observers having no information, classical or quantum, about the system being erased;  by contrast, our extended Landauer's principle, Eq.~\ref{eq:main}, covers observers with classical or quantum information).

We can think of entanglement as a thermodynamic resource, a sort of very concentrated fuel: the consumption of one unit of entanglement can simultaneously 
erase a qubit and convert $kT \ln 2$ of heat into work, two tasks that would otherwise require the consumption of one bit of classical information each.
In other words, quantum information can be a thermodynamic resource twice as powerful as classical information.

\subsection*{Erasure in the context of reversible computation}

The published version of this manuscript includes a
brief note on the application of erasure in algorithms to make
computation more thermodynamically efficient (Figure~1 and
Supplementary Information, Section~V, of~\cite{lidia11}). In the following we
explain the relation of our work to established results on reversible
computation, in particular~\cite{bennett73,fredkin82,bennett82, watrous08,bennett97}.

Consider a quantum algorithm with input $X$ and output $Y$ (the
algorithm may realize an arbitrary, not necessarily classical,
mapping). Using extra (initialized) ancilla registers, $R$, the
algorithm can always be implemented
reversibly~\cite{bennett73,fredkin82,bennett82}, corresponding to an isometry
that maps any initial state on $X$ to a joint state on $Y$ and
$R$. Eq.~\ref{eq:main} now tells us that the ancillas $R$ can in
principle be
erased (i.e., reset to their initial state) at a work cost rate of $\tilde{w}(R|Y) =
H(R|Y)\ k T \ln 2$. In general, $Y$ and $R$ may be
entangled and the work cost may be negative, so that erasing the
ancillas results in a gain of work.

Efficient erasure of the ancillas is well-established in theory of computation  for the case of deterministic algorithms (see, e.g., Figure~8 of~\cite{watrous08}).
First note that all deterministic functions can be made injective (so that $X$ is determined by $Y$) by treating the input as part of the output.
This makes the algorithm reversible and the entropy $H(R|Y)$ zero. Hence, there must exist a procedure for erasing the ancillas at no energy cost. This erasure can be done efficiently as follows.
After the execution of the algorithm, the classical output $Y$ is copied to a separate register. Then the reversible algorithm is run backwards, thereby resetting the ancillas $R$ to their initial state. Note that this
procedure requires the output $Y$ to be classical (otherwise the copy
operation may affect the joint state of $Y$ and $R$).

In fact, any probabilistic algorithm for a decision problem (or, more generally, the computation of a classical function, such as factoring) that receives a classical input can be boosted  to a virtually deterministic one by repeated iterations of the algorithm followed by a majority vote, so that the above considerations apply~\cite{bennett97,watrous08}.

However, the described procedures require both the input and the output of the algorithm to be classical.
It would be interesting to apply our results to the more general case of algorithms with quantum input or output. Examples could be the simulation of a physical system, or a tomography-type procedure that takes a finite number of copies of a quantum state as input and should output an estimation of its density matrix.

It is perhaps worth noting that our result only implies the existence of an erasure procedure with a given work cost or gain; we do not show how to implement such a procedure, or whether it can be done efficiently. %, which would require a case-by-case analysis.

%..................... BIBLIOGRAPHY .........................
%\bibliographystyle{plain}

\bibliographystyle{naturemag}

\bibliography{thermodynamics}

\end{document}